\documentclass[draft]{agujournal2019}
\usepackage{url} 
\usepackage{lineno}
\usepackage[inline]{trackchanges} 
\usepackage{soul}
\usepackage{amssymb}
\draftfalse

\journalname{JGR: Planets}

\begin{document}

%
%

\title{The nature and origins of sub-Neptune size planets}

%
%




\authors{Jacob L.\ Bean\affil{1}, Sean N.\ Raymond\affil{2}, \& James E.\ Owen\affil{3}}


\affiliation{1}{Department of Astronomy \& Astrophysics, University of Chicago, 5640 South Ellis Avenue, Chicago, IL 60637, USA}
\affiliation{2}{Laboratoire d’Astrophysique de Bordeaux, CNRS and Universit{\'e} de Bordeaux, All{\'e}e Geoffroy St. Hilaire, F-33165 Pessac, France}
\affiliation{3}{Astrophysics Group, Department of Physics, Imperial College London, Prince Consort Rd, London, SW7 2AZ, UK}





\correspondingauthor{J.\ L.\ Bean}{jbean@astro.uchicago.edu}




\begin{keypoints}
\item Sub-Neptune planets are rocky bodies that bifurcate into two classes based on their retention or loss of hydrogen-dominated atmospheres.
\item Sub-Neptune planets formed within gas-dominated disks from solids that experienced large-scale inward movement.
\item Atmospheric characterization of sub-Neptune planets has been frustrated by the presence of aerosols.
\end{keypoints}

%
%

%
%


\begin{abstract}
Planets intermediate in size between the Earth and Neptune, and orbiting closer to their host stars than Mercury does the Sun, are the most common type of planet revealed by exoplanet surveys over the last quarter century. Results from NASA's \textit{Kepler} mission have revealed a bimodality in the radius distribution of these objects, with a relative underabundance of planets between 1.5 and 2.0\,$R_{\oplus}$. This bimodality suggests that sub-Neptunes are mostly rocky planets that were born with primary atmospheres a few percent by mass accreted from the protoplanetary nebula. Planets above the radius gap were able to retain their atmospheres (``gas-rich super-Earths''), while planets below the radius gap lost their atmospheres and are stripped cores (``true super-Earths''). The mechanism that drives atmospheric loss for these planets remains an outstanding question, with photoevaporation and core-powered mass loss being the prime candidates. As with the mass-loss mechanism, there are two contenders for the origins of the solids in sub-Neptune planets: the migration model involves the growth and migration of embryos from beyond the ice line, while the drift model involves inward-drifting pebbles that coagulate to form planets close-in. Atmospheric studies have the potential to break degeneracies in interior structure models and place additional constraints on the origins of these planets. However, most atmospheric characterization efforts have been confounded by aerosols. Observations with upcoming facilities are expected to finally reveal the atmospheric compositions of these worlds, which are arguably the first fundamentally new type of planetary object identified from the study of exoplanets.

\end{abstract}

\section*{Plain Language Summary}
Planets with radii between that of the Earth and Neptune have been found around other stars in large numbers. It wasn't immediately obvious after their initial discovery what the basic characteristics of these planets are and how they formed because there aren't exact analogues of them in the solar system. Scientists have recently concluded that they are most likely Earth-like in composition based on measurements of how common objects of different sizes and densities in this regime are. However, there are two classes of these objects. The class of slightly larger objects harbors moderately thick atmospheres composed primarily of hydrogen gas. The other class of smaller objects are thought to have been born with similar atmospheres, but lost them during their subsequent evolution. Both classes of these planets must have formed very soon after the formation of their host stars in order to have started with hydrogen-dominated atmospheres, but the exact sequence of events leading to the birth of these objects remains uncertain. Efforts to directly study the atmospheres of these objects have been mostly stymied by heavy cloud layers. Observations with new telescopes are expected to yield detailed information on the atmospheres to further our understanding of these objects.

\section{Introduction} \label{sec:intro}
\begin{figure}
\noindent\includegraphics[width=\textwidth]{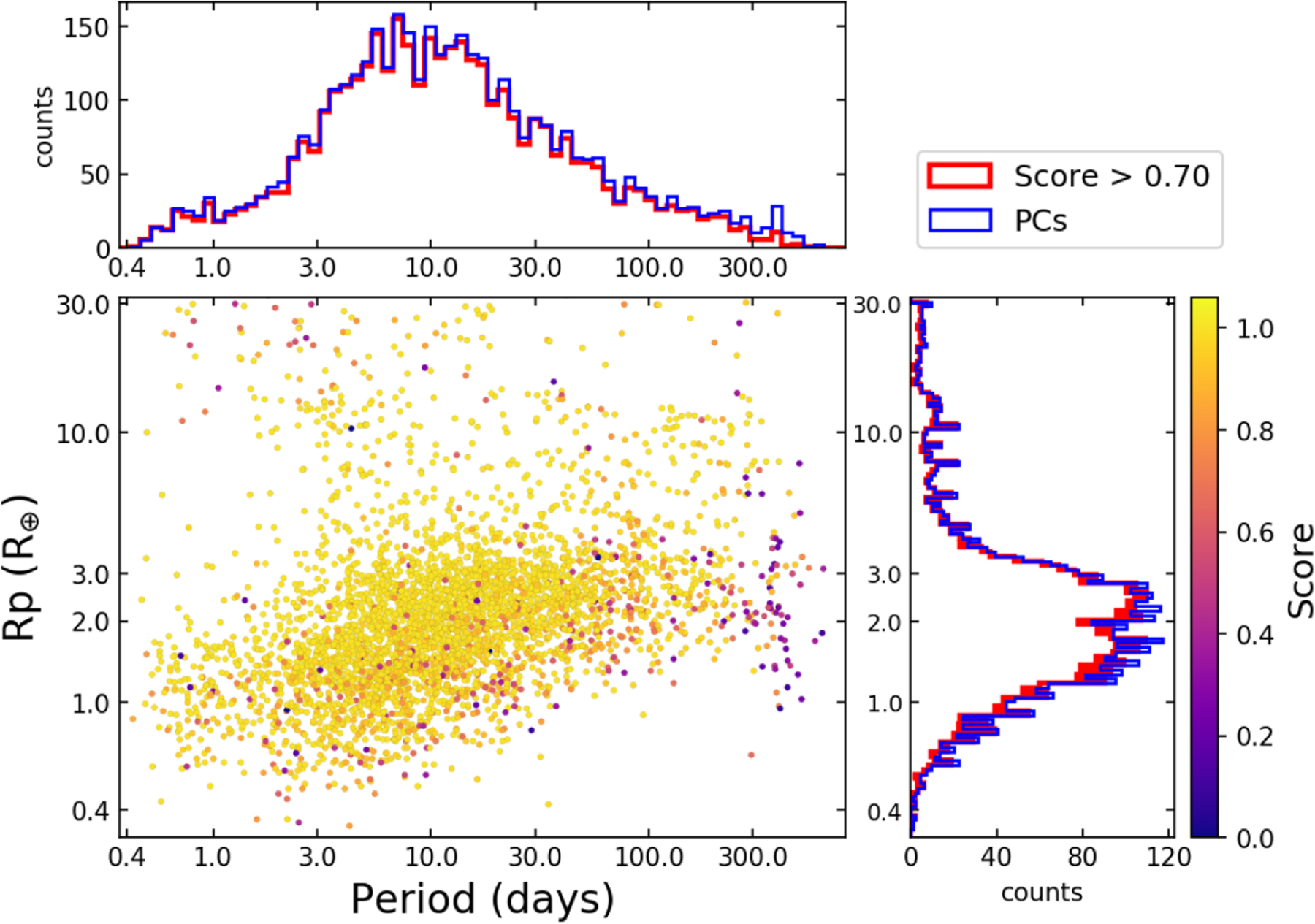}
\caption{Radii and orbital period for transiting planet candidates detected by \textit{Kepler} from \citeA{thompson18}. The color of the points (the ``Score'') is related to the estimated likelihood of being a true planet, where larger values indicate a higher likelihood. Most of the planet candidates that \textit{Kepler} discovered are smaller than Neptune and are likely to be real planets. The area in the bottom right of the figure is mostly empty due to selection effects. See \citeA{winn18} for more on selection biases of exoplanet surveys.}
\label{fig:kepler}
\end{figure}

The discovery that more than half of all Sun-like stars host close-in planets intermediate in size between the Earth and Neptune (``sub-Neptune size planets'') is perhaps the most profound discovery from NASA's \textit{Kepler} mission -- see Figure~\ref{fig:kepler} \cite{howard12,fressin13}. The existence of these planets in large numbers wasn't predicted by planet formation theories \cite{ida04,mordasini09}, and their provenance remains hotly debated \cite<e.g.,>[]{raymond08, rogers11, chiang13, hansen13, lee14, raymond14, schlichting14, lee16, ginzburg16, izidoro17, raymond18, adams20}. 

In this article we review the current state of knowledge of close-in, sub-Neptune size planets. We discuss the early history of the discovery and nomenclature of these objects in \S\ref{sec:history}. The most important insights to the nature of sub-Neptune planets have come from the statistical distribution of their radii and densities, which we review in \S\ref{sec:statistics}. We discuss the formation pathways implied by these results in \S\ref{sec:formation}. We summarize the results of atmospheric studies of these objects in \S\ref{sec:atmospheres}, and we conclude with a look at the future directions for research in this topic in \S\ref{sec:future}.



\section{Historical Perspective on the Characterization of Sub-Neptune Size Exoplanets} \label{sec:history}
A major hindrance to understanding the origins of sub-Neptune exoplanets has been the persistent question of their basic nature. Although this class of planet is essentially defined as being unlike previously known objects, it was natural to try to identify the best solar system analogue and associated formation history for comparison. The first examples of these objects that were discovered by Doppler (i.e., radial velocity) and transit surveys were nearly universally dubbed ``super-Earths'' \cite{rivera05, seager07, valencia07a, udry07, leger09, queloz09, charbonneau09, howard09, dawson10, winn11, demory11}. This naming was somewhat aspirational, reflecting the hopes that exoplanet detection had crossed a divide from gaseous to terrestrial, and that addressing questions of extrasolar habitability were just around the corner. However, only the minimum masses of the Doppler planets were known, thus the terrestrial nature of these planets couldn't be definitively established. Furthermore, the densities measured for some of the transiting planets were too low for them to be purely rocky bodies.

The early low-density, sub-Neptune size planets and the many planets that were discovered in the first data from \textit{Kepler} \cite{borucki11} prompted some to propose that these objects were best understood not as super-Earths, but instead as scaled down versions of the solar system ice giants, so-called ``mini-Neptunes'' \cite{barnes09, rogers10b, rogers11, marcy14, rogers15}. This term is used inconsistently in the literature. Sometimes it is used to mean a planet with a hydrogen-rich envelope, without regard for the composition of the core (i.e., ice/rock admixtures or rock only), while sometimes it is used to also invoke an ice-rich interior comparable to Uranus and Neptune \cite{nettelmann13, helled20}.

A third possibility for the nature of these planets was anticipated before any were discovered. \citeA{kuchner03} and \citeA{leger04} posited the existence of ``ocean-planets'' or ``water worlds'' (the term we will use), which are roughly defined as small planets ($M_{p}\,\lesssim\,10\,M_{\oplus}$) with significant water content (water mass fractions $\gtrsim\,10\%$) and lacking hydrogen-dominated atmospheres. A number of studies explored this idea following the early detections cited above \cite{valencia07a, fortney07, sotin07, adams08, rogers10b, fu10}.

The question of the fundamental nature of sub-Neptune exoplanets has persisted as we have moved into an era of precise density measurements for a large sample transiting planets \cite<e.g.,>[]{otegi20} because these planets exist in a part of the mass-radius parameter space that is a nexus of interior structure modeling degeneracy. Since these planets could be made of rock, ices, and/or gas, there are often multiple combinations of these materials that can match mass and radius data, and only the densest objects can be assured to be terrestrial \cite{selsis07, sotin07, valencia07b, adams08, rogers10a, nettelmann11}.

\begin{figure}
\noindent\includegraphics[width=\textwidth]{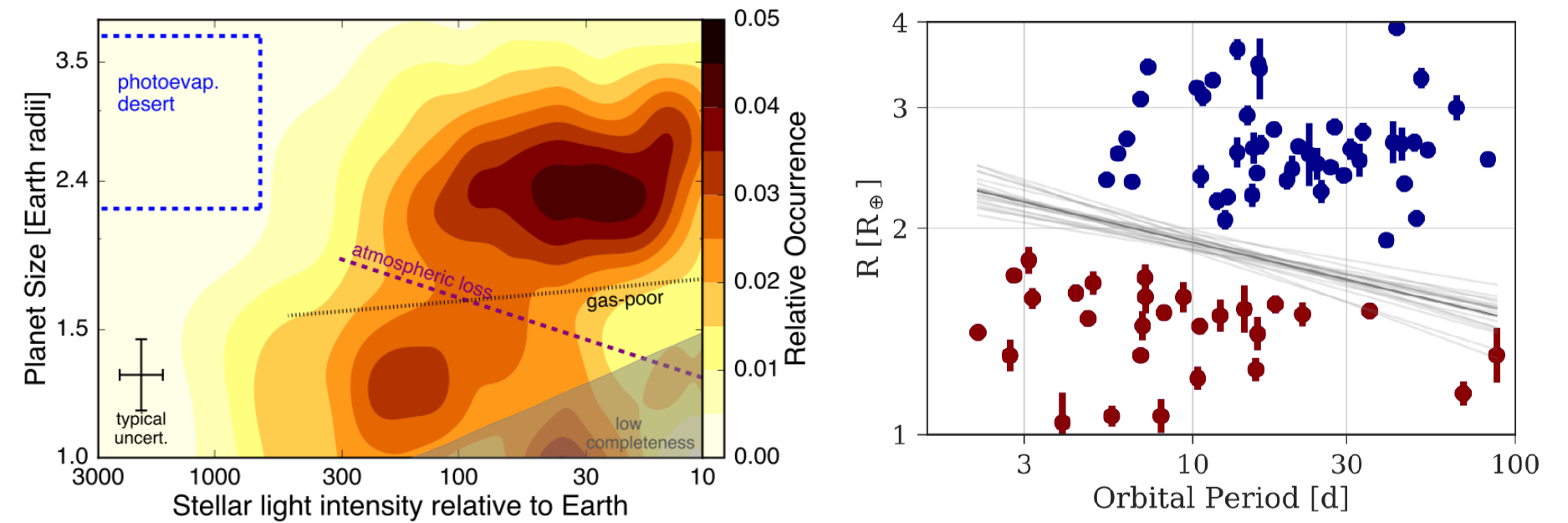}
\caption{Two views of \textit{Kepler's} planet radius gap for sub-Neptune size planets. In both cases the slope of the radius gap to smaller instellations and larger orbital periods matches the expectations of atmospheric mass loss being the key discriminant between the larger and smaller populations. \textit{Left:} Planet occurrence as a function of planet size and instellation with host star radii derived from spectroscopy and distances. Two peaks in the distribution centered at 2.4 and 1.3\,$R_{\oplus}$ are visible in the data. The two lines represent expectations for models of the formation and evolution of these objects from \citeA{lopez13}. Figure taken from \citeA{fulton17}. \textit{Right:} Precisely measured radii for individual planets from stellar characterization via asteroseismology. The lines present the best fit to the radius gap. This figure was originally presented as Figure 5 in \citeA{vaneylen18}.}
\label{fig:occurence}
\end{figure}

\section{Clues from Recent Population Studies} \label{sec:statistics}
Two recent observational results have shed new light on the question of the nature of sub-Neptune exoplanets. The first is the identification of a drop in the frequency of planets between 1.5 and 2.0\,$R_{\oplus}$ in the planet radius distribution using data from the \textit{Kepler} mission -- see Figure~\ref{fig:occurence} \cite{fulton17, fulton18, vaneylen18}. \textit{Kepler} used the transit technique to discover thousands of planets smaller than Neptune around Sun-like stars, thus enabling the first precise statistics of the frequency of close-in planets down to the size of the Earth. The primary observable from the transit technique is the ratio of the planet and host star radii \cite<although note that the stellar density can also be constrained directly from transit data;>[]{seager03}, thus knowledge of the host star properties are needed to convert \textit{Kepler's} measured transit depths into planetary radii. The host star properties of \textit{Kepler's} planets were estimated from photometry in the early days of the mission, with errors on the stellar radii of approximately 40\% \cite{brown11}. More precise stellar characterization from a combination of spectroscopy using ground-based telescopes \cite{johnson17}, distances from ESA's \textit{Gaia} mission \cite{fulton18}, and asteroseismology using \textit{Kepler's} time-series photometry \cite{vaneylen18} reduced the uncertainties on planetary radii to 5\% or less in the last few years.

The more precise measurements of \textit{Kepler} planetary radii from improved host star characterization revealed a bimodality in the radius distribution of sub-Neptune planets. Importantly, the gap, or valley, between the two peaks in the distribution has a dependency on orbital period, which has been interpreted as a trend in the incident stellar irradiation received by the planets (``instellation''). The slope of the gap with instellation matches the predictions of models where the two populations of planets both originally formed with hydrogen-dominated envelopes, but the more highly irradiated objects subsequently loss their envelopes (see the dashed line in the left panel of Figure~\ref{fig:occurence}). For the Sun-like stars that \textit{Kepler} primarily observed, gas poor formation models are ruled out because they predict the opposite slope in the radius gap from what is observed (see the dotted line in the left panel of Figure~\ref{fig:occurence}). Based on these and other similar results, it is widely held that atmospheric mass loss is the key process that sculpts the population of close-in sub-Neptune size planets orbiting Sun-like stars. However, the hypothesis that the 2 -- 4\,$R_{\oplus}$ planets are predominantly water worlds has not been totally abandoned, and we revisit this hypothesis at the end of this section.

Two main drivers for the mass loss that sculpts the sub-Neptune population have been proposed: ``photoevaporation'' \cite{owen13} and ``core-powered mass loss'' \cite{Ginzburg2018}. These two models share a similar physical basis: heating of the planet's upper atmosphere drives a hydrodynamic outflow, akin to a Parker wind \cite{Parker1958}, resulting in mass loss. However, the energy source that provides the heating of the upper atmosphere and drives the outflow differs. In the photoevaporation model, high-energy, ionizing, extreme ultraviolet (XUV) photons ($h\nu\sim$0.01-1~keV) produced in the stellar corona are absorbed by the planet's upper atmosphere. Due to the destruction of molecular coolants by the ionizing photons, the upper atmosphere it is heated to high temperatures (a few thousand to 10$^4$~K), driving a hyrodynamic outflow \cite{Lammer2003,Yelle2004,Garcia2007,MurrayClay2009,Owen2012}. In the core-powered mass-loss model, heating of the upper atmosphere from infrared (IR) radiation from the cooling planetary interior and bolometric irrdiation from the star similarly drives a hydrodynamic outflow, albeit a cooler and slower one. 

It is important to emphasise that the physical processes of XUV and IR/bolometric heating of the planet's upper atmosphere will both happen. However, what is not clear at this stage is which heating mechanism (and therefore whether photoevaporation or core-powered mass loss) dominates the mass loss from sub-Neptune sized planets. While both these heating processes are yet to be included self-consistently in a single model, we can hypothesize about the unified picture and speculate on its limits.

\begin{figure}
    \centering
    \includegraphics[width=\textwidth]{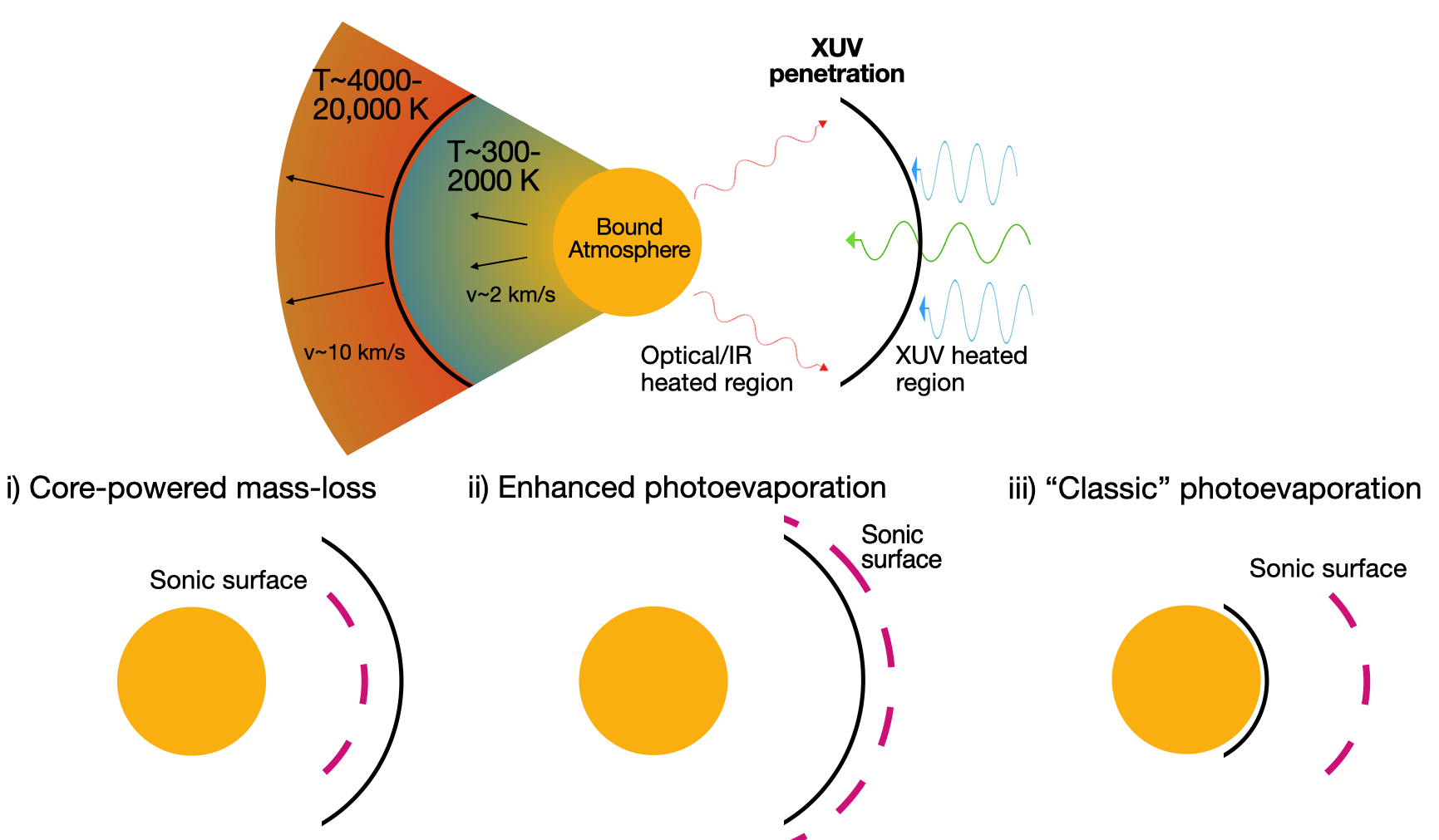}
    \caption{Schematic cartoon of the expected outflow structure in a unified picture for hydrodynamic mass loss from close in exoplanets (top) and the three (continuously connected) expected mass-loss regimes (bottom). The top panel shows the three layers to the planetary atmosphere. The bound atmosphere (yellow region) is where the hydrodynamic outflow is sub-dominant. The region heated by cooling radiation from the planetary interior (red photons) and the stellar bolometric luminosity (green photons) has an intermediate temperature (blue/green region). Finally, the region heated by stellar XUV irradiation (blue photons) is a few thousand Kelvin or more (orange region). The mass-loss regimes are shown from left to right as a function of increasing XUV luminosity (or decreasing cooling radiation). Core-powered mass loss occurs when the sonic surface sits interior to the penetration of XUV photons, which thus do not affect the outflow (i). When the sonic point occurs in the XUV heated region, but the cooling/bolometric heated region is not thin, photoevaporation is enhanced due to the larger sub-tended absorption area of the planet to XUV photons (ii), and finally when the cooling/bolmetric region is thin mass loss behaves as `classic' photoevaporation (iii). Only scenarios (i) and (iii) have been calculated, and only for each in isolation.}
    \label{fig:cartoon}
\end{figure}

The expected structure of the outflow is summarised in a schematic in Figure~\ref{fig:cartoon}, where due to the high cross-section for the absorption of XUV photons, they only penetrate into the very upper most layers of the atmosphere. Since the outflow upstream of the sonic surface is not in causal contact with the planetary atmosphere, the position of the sonic surface compared to the penetration depth of XUV photons essentially sets whether core-powered mass loss or photoevaporation dominates. We can imagine as the XUV luminosity increases (or the cooling radiation decreases) we transition from a regime where core-powered mass loss dominates, to one where the sonic surface occurs just inside the XUV heated region. At this point the cooling/bolometrically heated outflow is not thin (compared to the planetary radius) and increases the planet's effective cross-sectional area to the absorption of XUV photons. The increase in absorption of XUV photons could lead to mass-loss rates enhanced above the standard expectation for photoevaporation. Eventually, as the XUV luminosity increases, the upper atmosphere will be entirely XUV dominated and we return to the standard picture of photoevaporation.   

\begin{figure}
    \centering
    \includegraphics[width=0.6\textwidth]{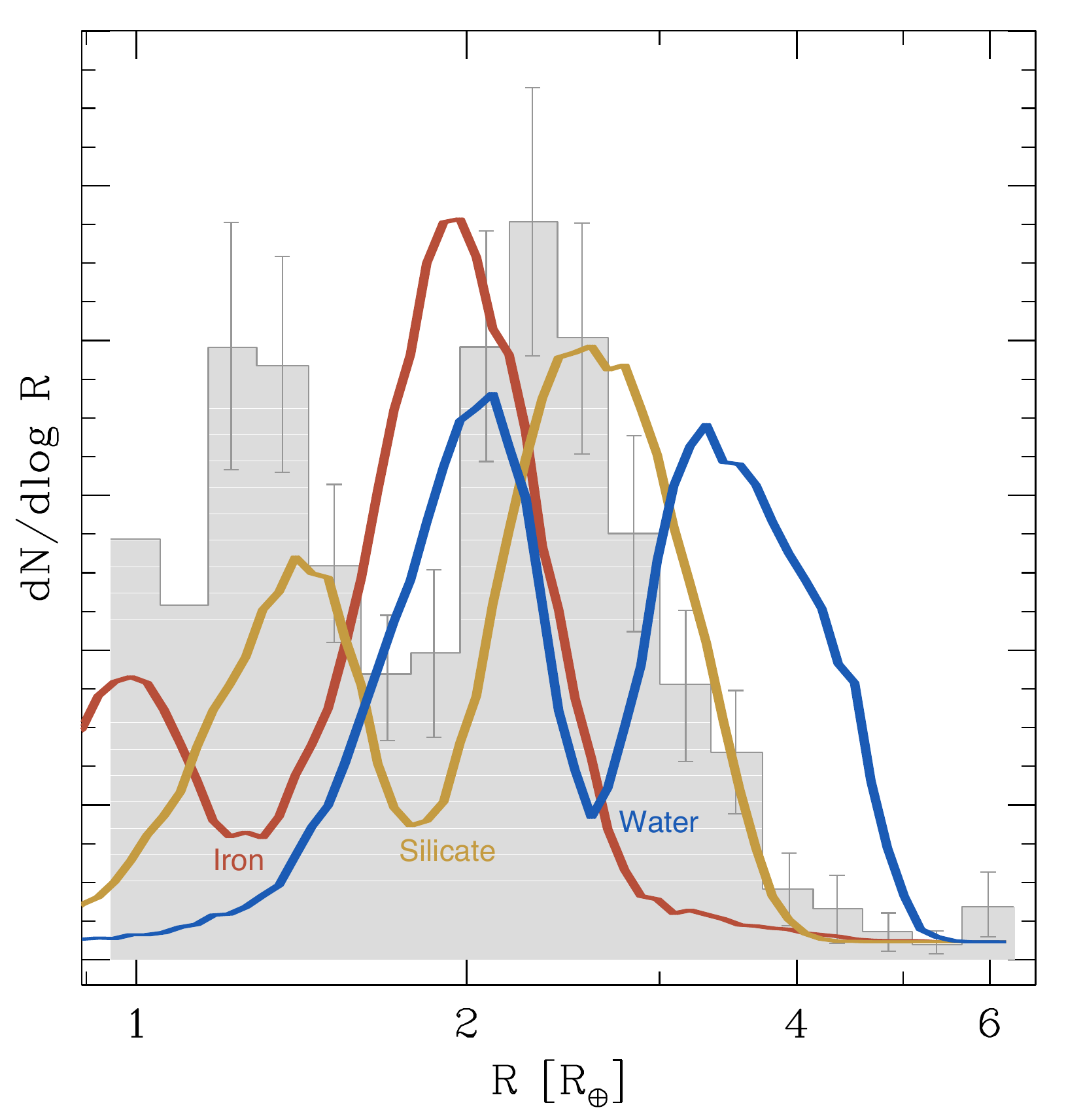}
    \caption{The observed radius distribution of {\it Kepler} planets with orbital periods $<$100\,days is shown as the grey histogram. The radius distributions predicted by the photoevaporation for different solid core compositions are shown as the colored lines. Lower density cores predict the radius gap to appear at higher radii. The observed radius distribution implies cores have densities consistent with an Earth-like rock-iron mixture (i.e., a model intermediate between the red and yellow models). More sophisticated models tightly constrain the silicate-to-iron ratio to be $\sim$3:1, i.e. consistent with Earth's composition \cite{RogersOwen2020}. Figure from \citeA{owen17}.}
    \label{fig:valley_comp}
\end{figure}

Both photovepoaration and core-powered mass loss suggests a unified explanation for the close-in sub-Neptune size planets: they are large (i.e., the mean mass is somewhere in the neighborhood of 3 -- 8\,$M_{\oplus}$) terrestrial planets born with hydrogen-dominated atmospheres that are a few percent by mass. Planets below the radius gap were fully stripped of these primordial atmospheres, while planets above the gap held on to their hydrogen-dominated atmospheres. Since the radius at which this transition occurs has been observed, and mass loss is sensitive to the planet's mass (with more massive planets better able to hold onto their natal hydrogen atmospheres), then the position of the radius gap is a probe of the core's composition. More volatile-rich cores are expected to have a radius gap at larger radii (see Figure~\ref{fig:valley_comp}), and this appears to be ruled out by the data.

Numerous works constrain the core compositions to be volatile poor and consistent with an Earth-like silicate-to-iron ratio \cite{owen17,Owen2019,wu19,Ginzburg2018,Gupta2019}. This conclusion appears to be robust, and is independent of whether the photoevaporation or core-powered mass-loss model is used. Further, the fact that the radius gap is observed to be a relatively sharp feature indicates that there is not a large spread in the core densities. Recent statistical fits of the radius gap within the framework of the photoevaporation model suggest the mean density of an Earth-mass core is $\rho_{1{\rm M}_\oplus}=5.1\pm0.4$~g~cm$^{-3}$, with a variance in the density of Earth-mass cores at the $< 1$~g~cm$^{-3}$  level. This mean core density implies, for a typical core mass of $6~$M$_\oplus$, the water content can be no higher than $20\%$ and this is even in the hypothetical and unlikely case that it's composed of iron and water only, with no silicates \cite{RogersOwen2020}.  While photoevaporation and core-powered mass-loss models agree on the core composition, they diverge on other inferred properties. In particular, the photoevaporation model suggests a positive linear correlation between core mass and stellar mass \cite{wu19}, while core-powered mass loss implies the core mass is independent of stellar mass, although some correlation is not ruled out \cite{Gupta2020}.

\begin{figure}
    \centering
    \includegraphics[width=0.7\textwidth]{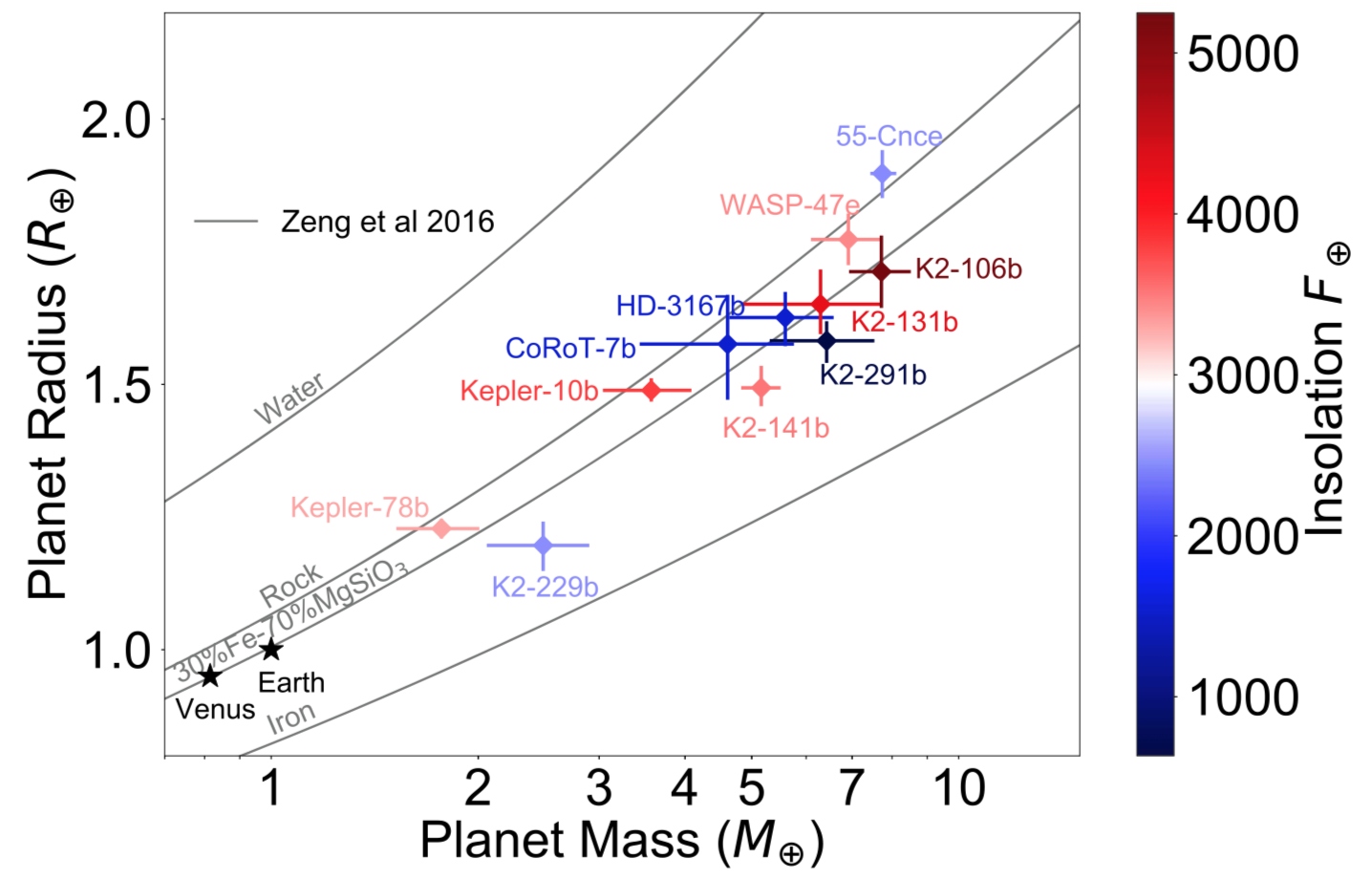}
    \caption{Radius vs.\ mass from a uniform analysis of small, highly-irradiated planets. These planets should not have substantial gaseous envelopes, thus removing a degree of freedom from interior structure models. The data are tightly clustered around the Earth-like composition line, suggesting a common composition for rocky planets. The higher density outlier K2-229b could have a higher iron fraction from collisions \cite<although note that current models struggle to create very iron-rich planets from collisions;>{scora20}, while the low density outlier 55\,Cnc\,e could be the rare small planet with a significant volatile content or no core. Figure from \citeA{dai19}.}
    \label{fig:hot_Earths}
\end{figure}

A second key observational result comes from the careful analysis of the masses and radii of ultra-short period planets -- see Figure~\ref{fig:hot_Earths} \cite{dai19}. These planets are so highly irradiated that it is very unlikely that they harbor hydrogen-dominated atmospheres, thus eliminating a degree of freedom in the interior structure models. \citeA{dai19} found that most of the small ultra-short period planets were consistent with an Earth-like terrestrial composition, with the exception of two out of the 11 planet sample that were more or less dense. The low density object, 55\,Cnc\,e (see also a discussion of its atmosphere in \S\ref{sec:atmospheres}), is the most interesting in the context of this review because its large radius suggests it either has a significant component of low-density volatiles, or it has no core and is predominantly made up of Ca and Al minerals that condense at high temperatures \cite{dorn19}. The potential presence of a significant amount of volatiles is in tension with the statistics of the radius gap, which can be fully explained with a population model that has essentially no water-rich planets \cite{RogersOwen2020}. 55\,Cnc\,e is also part of a system that is unusually rich in giant planets \cite{fischer08}, so it might not be representative of the broader sub-Neptune planet class.


As mentioned above, the mass-loss hypothesis for the nature and origins of sub-Neptune planets currently has the most traction in the field, but alternative hypotheses have not been fully abandoned. In particular, \citeA{zeng19} and \citeA{mousis20} have extended earlier work on water worlds and shown that the population of 2 -- 4\,$R_{\oplus}$ planets can be explained by internal structure models with a large fraction of volatiles. They have shown that roughly 50/50 admixtures of rock and water can reproduce the 2.5\,$R_{\oplus}$ peak in the \textit{Kepler} radius distribution, with variations in water abundance explaining the range of sizes and densities for these planets instead of variations in the hydrogen fraction. The water world model has not gained as wide acceptance as the mass-loss model in the exoplanet community mainly because it doesn't explain the two key observed correlations with instellation that are described above (i.e., the radius distribution with orbital period and the densities of highly irradiated planets). While it remains true that models for the internal structures of \textit{individual} planets in the 2 -- 4\,$R_{\oplus}$ regime are degenerate, and thus water-world solutions are possible, the mass-loss hypothesis offers a simple, unified explanation for the trends in the entire \textit{population} of sub-Neptune planets. Nevertheless, work should continue to test and refine all plausible models.


\section{Implied Formation Pathways} \label{sec:formation}
Understanding the origins of systems of close-in sub-Neptunes requires a change in our frame of reference. The majority of planet formation studies to date have focused on our solar system~\cite<e.g.>{safronov72,wetherill78}. This is not surprising given that exoplanets were only discovered in the past few decades, whereas the origins of the solar system planets have been pondered for centuries.  Close-in planetary systems represent fundamentally different outcomes of planet formation that must be more common than the one that produced the solar system. While the same physical processes should govern the formation of all planetary systems, the specific sequence of events must play a key role in shaping their orbital architectures~\cite<see discussion in>[]{raymond18b}.  

\begin{figure}
\noindent\includegraphics[width=\textwidth]{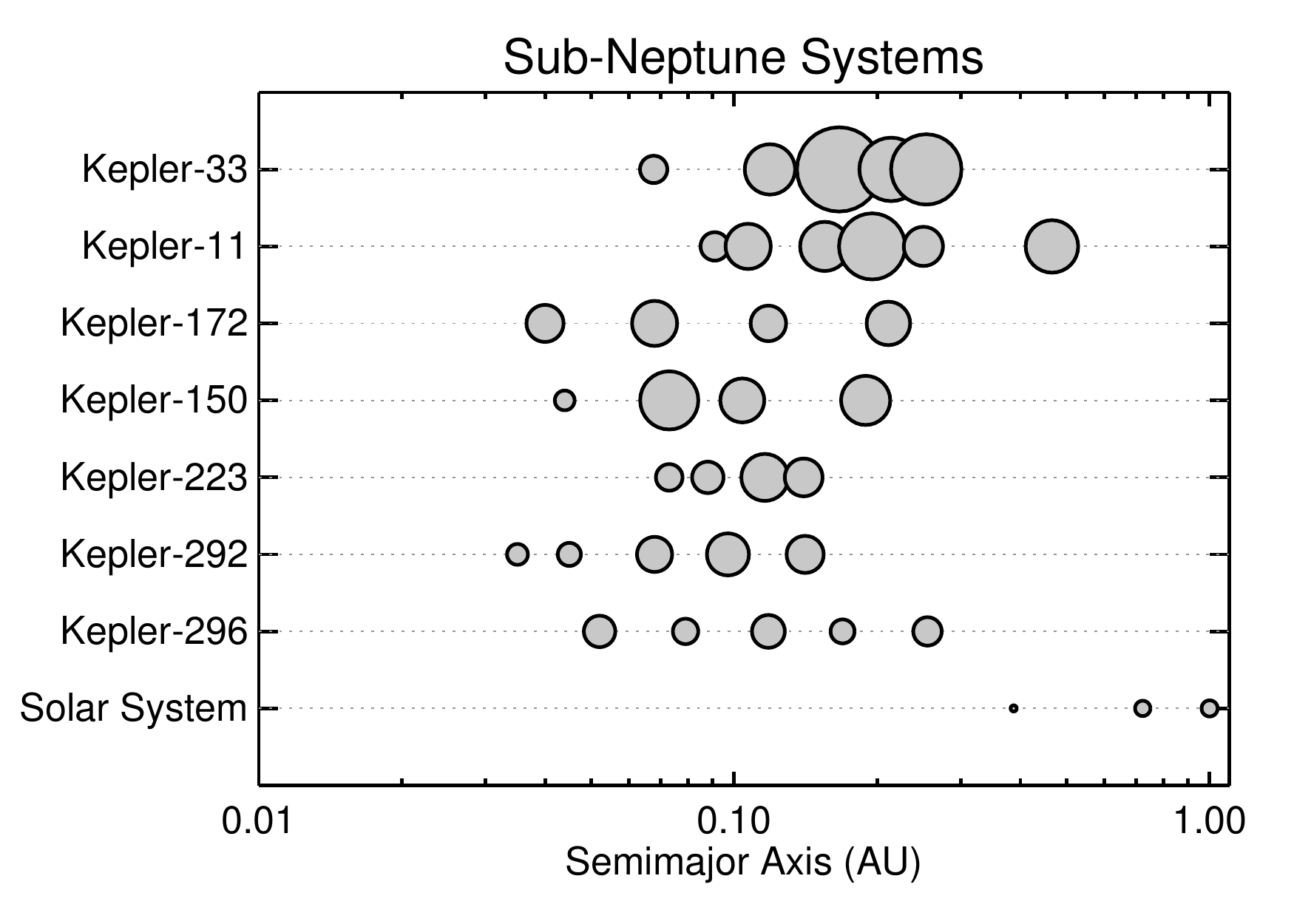}
\caption{Orbital architectures of seven selected sub-Neptune systems.  Each system contains at least one sub-Neptune planet with $2 R_\oplus < R_{p} < 4 R_\oplus$, and systems are ordered vertically by the median planet size, from the largest (top) to smallest (bottom). The size of each planet is proportional to its measured physical size. The $x$ axis is logarithmic such that the distance between neighboring planets in a given system is a measure of their orbital period ratio. For scale, the Mercury-Venus orbital period ratio is 2.55 and the Earth-Venus ratio is 1.63. Roughly half of Sun-like stars have rich systems of close-in sub-Neptune size planets like those shown here.}
\label{fig:arch}
\end{figure}

Observational constraints on formation models come from the bulk properties of the observed population of close-in sub-Neptunes.  Specific quantities include these planets' sizes, densities and orbital distances~\cite<including the correlated planet sizes within individual systems;>[]{ciardi13,wu13,weiss18,weiss20}. Given that many sub-Neptunes are found in multiple-planet systems (see Figure~\ref{fig:arch}), other constraints come from the multiplicity distribution~\cite{youdin11,fang12,tremaine12,johansen12} of observed transiting systems (in other words, how many planets transit each star), as well as the orbital period ratio distribution of adjacent planets~\cite{lissauer11b,fabrycky14}.

At least seven models exist to explain the origins of close-in sub-Neptunes~\cite{raymond08}, most of which were proposed prior the launch of the {\em Kepler} space telescope. Some of these models~\cite<e.g.>[]{zhou05,fogg05,raymond06} relied on the dynamical influence of giant planets and can be ruled out on the simple grounds that the measured occurrence rate of super-Earths is far higher than that of gas giants~\cite{mayor11,howard10,howard12,fressin13}. The very simplest model -- often called ``in-situ accretion'' -- proposed that close-in planets grew in the same way as our own terrestrial planets, by successive impacts between ever-larger planetary embryos within disks that were massive enough to have many Earth masses of solids very close to their stars~\cite{chiang13}. That model can also be ruled out because it assumes that the planets grew in place, close to their current orbital distances.  As such, the model is not self-consistent: any disk massive enough to form such planets in-situ would drive orbital migration at such a fast rate as to make migration a central process~\cite{bolmont14,inamdar15,ogihara15}. 

\begin{figure}
\noindent\includegraphics[width=\textwidth]{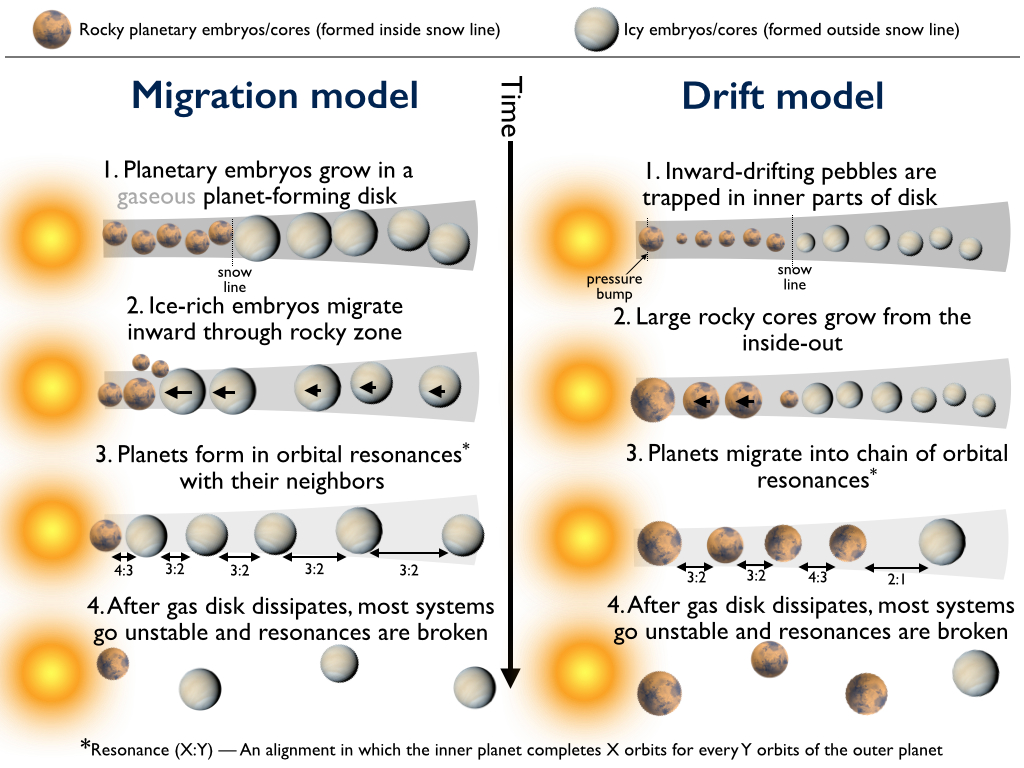}
\caption{Conceptual formation pathways for close-in low-mass planets. }
\label{fig:formation}
\end{figure}

There are currently two plausible models to explain the origins of close-in small planets (see Figure~\ref{fig:formation}). Both invoke large-scale inward movement of solids within gas-dominated planet-forming disks but at very different size scales. In the {\em drift} model the majority of growth takes place close in, from mass that has drifted inward. In the current paradigm of planet formation, dust grains coagulate and grow until they become large enough to partially decouple from the gas and drift inward~\cite<see>[]{ormel17,johansen17}.  Dust particles that grow big enough to drift rapidly are often referred to as ``pebbles''. Observations of gas-rich disks around young stars commonly find evidence for the existence of pebbles~\cite{natta07,perez15}, and it is inferred that they drift inward because dust disks are observed to be more compact than gas disks~\cite{andrews12,cleeves16,trapman19}. The ring-like structures observed in many disks~\cite{alma15,andrews18} are thought to be produced by growing and drifting dust/pebbles~\cite<e.g.>[]{dullemond18,birnstiel18}. Drifting pebbles may be trapped at a pressure bump in the inner parts of the disk~\cite{chatterjee14,chatterjee15,boley14,hu16,hu17,jankovic19}. Indeed, MHD simulations of the inner regions of disks find that pressure bumps should exist close in and are capable of trapping drifting particles~\cite{flock17,flock19}. The next phases of growth are thought to involve gravitational instability to form planetary embryos, followed by mutual collisions and orbital migration~\cite<e.g., see>[]{hansen12,hansen13,dawson15,moriarty16,flock19}. 

In the {\em migration} scenario mass is delivered to the inner disk in the form of large planetary cores rather than drifting pebbles. Cores are assumed to form across the disk by planetesimal and pebble accretion~\cite<e.g.>[]{johansen17}. Massive cores likely form preferentially at or past the snow line, where pebble accretion is accelerated~\cite{lambrechts14,morby15,ormel17}. Once they reach a critical mass, cores migrate inward until they reach the inner edge of the disk~\cite<e.g.>[]{terquem07,ogihara09,mcneil10,ida10,rogers11,cossou14,coleman16}. As in the drift model, the final phases of growth involve giant collisions between cores. It is worth noting that our understanding of migration is incomplete, as even the highest resolution simulations to date cannot fully resolve the behavior of low-mass planets in low-viscosity disks~\cite{mcnally19}.

In each scenario, the growing planets accrete gaseous envelopes directly from the disk. The structure of these primordial atmospheres is determined by a complex competition between gas flow within the disk, thermal evolution, and loss during impacts \cite{ikoma12,lee14,lee16,schlichting14,inamdar16,ginzburg16,coleman17,lambrechts17,lambrechts19}. Once the gas disk has dissipated, these atmospheres are subject to loss processes (see \S\ref{sec:statistics}).

The {\em drift} and {\em migration} models predict different compositions for close-in planets.  Pebbles should lose their volatiles as they drift inward across the snow line~\cite<which is itself moving inward as the disk evolves; e.g.>{oka11,ida19} such that the planets formed in the {\em drift} model should be rocky, with little water.  In contrast, large migrating cores should retain the bulk of their volatiles. But exactly how water-rich should planets be in the {\em migration} scenario?  Unfortunately, this is currently unclear.  If planetesimals -- the seeds of super-Earths -- preferentially form just past the snow line~\cite{armitage16,drazkowska17,schoonenberg17} and grow further by pebble accretion (while migrating), then their bulk water contents are often a few to ten percent \cite<according to simulations;>[]{bitsch19}.  If, however, planetesimals form quickly across a broad swath of the disk then the final super-Earths are likely to be closer to 50\% water by mass \cite{izidoro19}.  Yet large migrating cores shepherd material interior to their orbits and catalyze the formation of even closer-in planets, which themselves are often volatile-poor~\cite{izidoro14,raymond18}. In simple terms, while the {\em drift} model predicts high-density volatile-poor planets, the {\em migration} model predicts a diversity of volatile contents of such planets, often within the same system. Very high water contents are at odds with the compositional inferences discussed above within the context of evaporative mass loss of the atmospheres of close-in planets (see \S\ref{sec:statistics}). Future high-precision measurements of the bulk densities of close-in planets, coupled with interior structure models, may distinguish between these model predictions.

The late dynamical evolution of the {\em drift} and {\em migration} models should be similar~\cite<see discussion in>[]{raymond18b}. In each model, massive planets form quickly and are close to the inner parts of the disk while the disk is still dense.  This implies that migration must be important during the later parts of gaseous disk phase. Migrating cores tend to form configurations in which each pair of neighboring planets is in mean motion resonance  \cite<e.g.>[]{terquem07,cresswell08}. In these ``resonant chains'', the innermost planet is anchored at the inner edge of the gaseous disk, which provides a positive torque to balance the negative ones felt by the other planets \cite{masset06,romanova19}. Resonant chains often become dynamically unstable after the gaseous disk dissipates, leading to phase of late giant impacts \cite{terquem07,ogihara09,cossou14}. This is the foundation of the {\em breaking the chains} model, which can match the period ratio and multiplicity distributions of {\it Kepler}'s close-in planets as long as 95\% or more of resonant chains become unstable \cite{izidoro17,izidoro19}. In this context, exotic multi-resonant systems such as TRAPPIST-1 \cite{gillon17} and Kepler-223 \cite{mills16} represent the rare resonant chains that remained stable after the disk dissipated. While the late stages of growth and migration of the {\em drift} model have not yet been modeled, we expect them to follow the {\em breaking the chains} pathway.

The distribution of sub-Neptunes' H/He atmospheres place constraints on formation models.  For example, envelopes containing a few percent of a planet's mass are needed to explain the radius valley (see \S\ref{sec:statistics}), yet it remains unclear why planets would not accrete substantially more gas from the disk. Several processes have been proposed to explain this, including delayed atmospheric cooling due to high opacity~\cite{lee14}, dissolution of $H_2$ in magma oceans~\cite{kite19}, and rapid disk photoevaporation~\cite{owen16,ginzburg16,ogihara20}. Giant impacts between growing sub-Neptunes are likely to lead to loss of the planets' primordial H/He atmospheres, especially for close-in young planets~\cite{biersteker19}. Yet if impacts are generic and systematically remove H/He envelopes, why do sub-Neptunes exist at all?  The answer to this apparent contradiction is not immediately clear. Perhaps impacts often occur before the full dissipation of the gaseous disk~\cite{esteves20} such that a thin atmosphere can still be accreted~\cite{lee15}.  Or perhaps resonant chains of planets are spread out by a different mechanism, such as by the {\em magnetospheric rebound} migration torque from the expanding disk cavity~\cite{liu17}. We expect future formation models to take advantage of detailed compositions constraints from studies of sub-Neptune atmospheres (see \S\ref{sec:future}).

How does the solar system fit into this picture? Why are there no super-Earths or sub-Neptunes close to the Sun?  A number of solutions to this very relevant problem have been proposed~\cite<for a detailed discussion, see>[]{raymond18b}. Some models propose that close-in planets did indeed form around the Sun but did not survive.  However, these scenarios are hard to reconcile with observations.  For example, if our Sun's close-in planets were collisionally ground to dust~\cite{volk15}, why are such planets so common around other stars and why are their orbital spacings suggestive of a late phase of giant impacts~\cite{pu15,izidoro17}?  Likewise, if planets formed close to the Sun but migrated away, either outward toward the giant planet region~\cite{raymond16}, or inward onto the Sun~\cite{batygin15}, then how can we reconcile this with the abundance of close-in exoplanets?

At present it seems more likely that some mechanism prevented the formation of close-in super-Earths or sub-Neptunes around the Sun. Perhaps Jupiter's growing core reached the pebble isolation mass~\cite{lambrechts14,bitsch18} and starved the inner solar system of inward-drifting pebbles, thus preventing the terrestrial planets from growing massive enough to migrate~\cite{lambrechts19}. This mechanism could explain the isotopic dichotomy of chondritic meteorites~\cite{warren11}, which appears to require spatial segregation of pebbles in the early solar system due to Jupiter's growing core~\cite{kruijer17,kruijer20} or perhaps pressure bumps in the disk~\cite{brasser20}. However, if the giant planets' cores grew fast enough to block the pebble flux into the inner solar system, why didn't they migrate farther inward themselves? Another possibility is that the full-grown Jupiter and Saturn blocked the inward migration of the progenitors of the ice giants~\cite{izidoro15,izidoro15b}. If that were the case, then we would expect to observe an anti-correlation between close-in planets and outer gas giants; that correlation is not presently observed~\cite{barbato18,zhu18,bryan19}.  Thus, while it remains an active area of research, it is unclear exactly which pieces of the puzzle are responsible for the lack of close-in large planets in the solar system.

\section{A Decade of Super-Earth Atmosphere Studies} \label{sec:atmospheres}
It has been recognized since the early days of sub-Neptune size exoplanet discovery that understanding their atmospheres holds a key to revealing the nature and origins of these mysterious objects. One reason for this is that determination of the atmospheric composition could help break the degeneracies in interior structure models and uniquely constrain their bulk makeup \cite{adams08, miller-ricci09, rogers10a, valencia13}. If the atmospheric compositions of these planets could be directly determined that would provide an important boundary condition for the models used to match the planets' masses and/or radii.

Another reason for the importance of these planets' atmosphere is that the composition of the atmosphere itself is also an important record of a planet's formation and evolution \cite{miller-ricci10, rogers10b, benneke13}. For example, both primary and secondary type atmospheres are possible for these planets, and these different types of envelopes can be distinguished by their composition. Furthermore, even primary atmospheres could be altered by interaction with the planet's interior, thus offering a potential view to the detailed composition of the bulk \cite{kite19,kite20}.

There are still very few direct observational constraints on the compositions of sub-Neptune size exoplanet atmospheres despite considerable effort over the last decade. Most efforts to date have focused on transmission spectroscopy observations because this is in principle the most efficient way to detect the atmospheres of these objects with existing facilities (as opposed to measurements of thermal emission or reflected light via secondary eclipses and phase curves). Transmission spectroscopy is also particularly well suited to addressing the key question of the hydrogen content of these planets' atmospheres because the size of spectral features in transmission spectroscopy measurements is primarily sensitive to the scale height of the atmosphere, which itself is mainly set by the abundance of hydrogen gas through its impact on the mean molecular weight \cite{miller-ricci09}.

Unfortunately, the vast majority of transmission spectroscopy measurements for sub-Neptune sized planets have yielded featureless, or so-called ``flat'' spectra \cite<e.g.,>[]{bean10b, bean11, desert11, berta12, fraine13, kreidberg14, knutson14, dewit16, diamond18, libby20, guo20}. The featureless spectra for the planets that must have gaseous envelopes (i.e., those with $R_{p}\,\gtrsim\,2\,R_{\oplus}$) can be explained by the presence of thick aerosols at high altitude obscuring our view of the bulk of the atmospheres, and thus they generally can't constrain the compositions of the planets' atmospheres. Aerosols are a particularly pernicious problem for super-Earths because these objects are already hard to observe due to their small size (reminder: signal sizes in transmission spectroscopy scale as $R_{p}^{3}$ all else being equal), and because these planets are typically cooler, which enhances aerosol formation \cite{miller-ricci12,morley13,morley15,hu14,mbarek16,kawashima19}. The featureless transmission spectra for planets with $R_{p}\,\lesssim\,2\,R_{\oplus}$ are consistent with these planets lacking cloudless, hydrogen-dominated atmospheres, but aren't constraining beyond that.

\begin{figure}
\noindent\includegraphics[width=\textwidth]{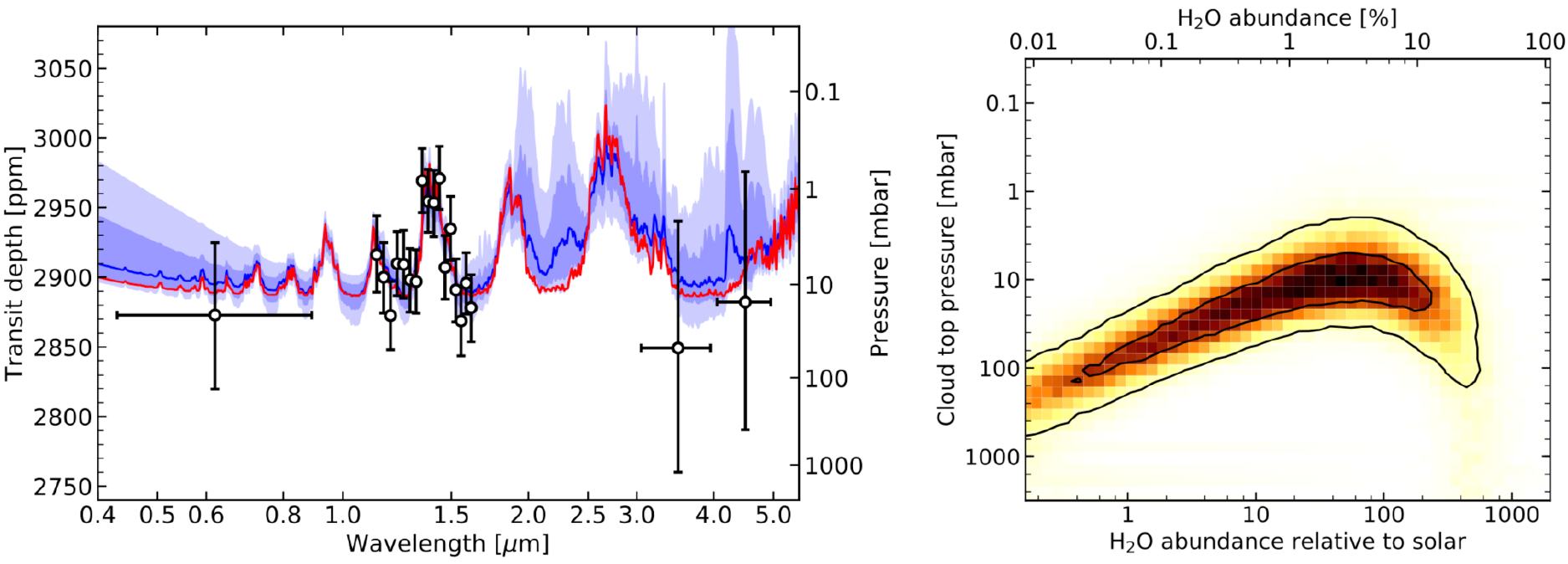}
\caption{\textit{Left:} The transmission spectrum of the sub-Neptune size planet K2-18b (black points with errors) compared to models (solid lines and shaded confidence intervals). The increase in transit depth detected near 1.4\,$\mu$m is identified as absorption due to water vapor. \textit{Right:} Derived constraints on the cloud top pressure and the water vapor abundance from a retrieval analysis of the transmission spectrum. The solid black lines represent the 1 and 2$\sigma$ confidence regions. Despite the degeneracy between the cloud top pressure and the water vapor abundance it is clear that the planet has a hydrogen-dominated atmosphere due to the detectability of the spectral features. Future observations with the \textit{James Webb Space Telescope} will be able to precisely constrain the atmospheric compositions of this and other similar planets and thus provide important constraints on their formation. Figures taken from \citeA{benneke19}.}
\label{fig:k2-18}
\end{figure}

There have been two notable successes in detecting features in the transmission spectra of sub-Neptune sized planets, both from \textit{Hubble Space Telescope} Wide Field Camera 3 (WFC3) observations. \citeA{tsiaras16} presented the detection of relatively large features in the transmission spectrum of 55\,Cnc\,e ($R_{p}\,=\,1.9\,R_{\oplus}$), and \citeA{benneke19} and \citeA{tsiaras19} present a much more convincing detection of features in the spectrum of K2-18b ($R_{p}\,=\,2.6\,R_{\oplus}$). Both of these detections imply hydrogen-dominated atmospheres but with large uncertainties in the overall heavy element abundance (metallicity) due to modeling degeneracies and the limited information content of the data. The presence of a hydrogen-dominated atmosphere on K2-18b (see Figure~\ref{fig:k2-18}) is consistent with the interpretation of the planet radius gap in the population statistics described above. However, the idea of a similar atmosphere on 55\,Cnc\,e is harder to reconcile given its very high level of irradiation (it has an instellation $S\,\sim\,2500\,S_{\oplus}$). The 55\,Cnc\,e data are also suspect because the host star is right at the brightness limit for WFC3, thus raising the possibility of unmitigated instrument systematics \cite{swain13,wilkins14,hilbert14}. In contrast, \citeA{jindal20} ruled out the presence of a hydrogen-dominated atmosphere containing water vapor for this planet using ground-based high-resolution spectroscopy.

55\,Cnc\,e is also one of the few sub-Neptune size planets that is amenable to thermal emission measurements. The \textit{Spitzer Space Telescope} full orbit phase curve for this planet has been shown to have a large amplitude, which is indicative of poor day-night heat redistribution (the planet is likely tidally locked), but also a large hot spot offset, which is indicative of substantial heat transport \cite{demory12,demory16b}. Surprisingly, the dayside thermal emission of the planet also shows variability, thus making the observations even more difficult to interpret \cite{demory16a}. Ultimately, 55\,Cnc was also at the brightness limit for the now defunct \textit{Spitzer}, and unrecognized instrument systematics could have impacted these measurements as well.

Beyond the attempts to directly observe the bulk atmospheres of sub-Neptune size planets, there have also been clever attempts to deduce the composition of these planets' atmospheres through observations of their thermospheres, exospheres, and winds. These observations have been obtained for neutral hydrogen at Lyman\,$\alpha$ \cite{ehrenreich12,bourrier17,waalkes19,dossantos20,garcia20} and in the helium infrared triplet \cite{kasper20}. The detection of neutral hydrogen in an escaping wind would constrain the mass-loss rate of the atmosphere but would have an ambiguous interpretation with regards to the composition because neutral hydrogen could be produced from the photodissociation of H$_{2}$ and H$_{2}$O. Helium is a more promising species from this standpoint because it would only exist in large quantities for a primary atmosphere that was accreted from the protoplanetary nebula. However, the most observable helium feature, the infrared triplet, arises from a metastable state that requires a finely tuned spectrum of UV irradiation from the host star to be populated \cite{oklopcic19}.

Unfortunately no clear detections of the upper atmospheres of super-Earths have been made so far. The most promising result is the tentative detection of neutral hydrogen for K2-18b based on a partial transit observed with \textit{Hubble} \cite{dossantos20}. This is consistent with the observation of the bulk atmosphere described above, but more data are needed to confirm the result.

\section{Future Directions} \label{sec:future}
Approximately 15 years since their first discovery, the nature and origins of sub-Neptune size exoplanets have started to come into focus. The global picture that has recently emerged from population level studies is that most close-in, sub-Neptune size planets are actually large terrestrial bodies, with the absence (``true super-Earths'') or presence (``gas-rich super-Earths'') of hydrogen-dominated atmospheres separating them into two classes. These objects are likely poor in volatiles ($\lesssim$10\% by mass), and their final assembly occurred close to their host stars in the presence of a gas-rich disk. It has been tempting to compare these objects to the solar system ice giants Uranus and Neptune because they have hydrogen-dominated atmospheres as a common factor \cite<e.g.,>{atreya20, wakeford20}. However, the more we learn about these objects the less appropriate this comparison is. Uranus and Neptune have roughly 10 times more hydrogen and helium by mass than the typical gas-rich super-Earth, their bulk and envelopes are likely rich in volatiles, and their formation histories must be quite different to have arrived at very different orbital distances.

The distinct internal structures of gas-rich super-Earths, i.e. rock overlaid by thick, hydrogen-dominated atmospheres, leads us to propose that these objects are the first fundamentally new type of planetary object identified from the study of exoplanets. There are a number of observations that can be done to test this hypothesis. One ongoing area of work is the precise measurement of masses and radii for sub-Neptune size planets orbiting stars with a range of masses and ages, and with a wide range of orbital separations. These observations should seek to determine how the planet radius gap varies with stellar mass \cite{cloutier20, hardegree20} and age \cite{berger2020}, and to ultimately reveal the statistical distribution of planet densities in the multi-dimensional parameter space. Early results on this topic from further analysis of \textit{Kepler}/\textit{K2} data have yielded tentative evidence that super-Earths form in gas poor disks around low-mass stars \cite{cloutier20}, and that the mass-loss timescale for these planets around stars with masses $\gtrsim$1\,M$_{\odot}$ is approximately a Gyr, which is a potential signpost to the core-powered mechanism \cite{berger2020}. Continuing work on this topic is currently enabled through the detection of transiting planets around bright stars by NASA's \textit{TESS} mission \cite<launched 2018;>{ricker18} and ESA's \textit{CHEOPS} mission \cite<launched 2019;>{benz20}, and will be furthered by ESA's \textit{PLATO} mission \cite<scheduled for launch in 2026;>{rauer14}.

Another key observation that can be done for sub-Neptune size planets is precise spectroscopy to reveal their atmospheric compositions. While such observations have been mostly stymied so far, the increased sensitivity and spectral range of the \textit{James Webb Space Telescope} \cite{beichman14, green16} and the next generation of ground-based Extremely Large Telescopes \cite{hood20,ghandi20} are expected deliver breakthroughs on this topic. Spectroscopy of gas-rich super-Earths should seek to determine if the metallicities of their atmospheres follow the trend of increasing metallicity with lower planet mass that is expected from extrapolating from giant planet formation \cite{fortney13}. These observations may also reveal atmospheric carbon-to-oxygen abundance ratios, which are a tracer of formation location and migration \cite{oberg11,madhusudhan14}. Gas-rich super-Earths are expected to have deep magma oceans in contact with their atmospheres, thus yielding unique chemistry in atmospheric gases that could be detectable \cite{kite20}, as well as potentially sculpting the populations statistics at large sizes \cite{kite19}. 

On the theoretical side, work combining models of photoevaporation and core-powered mass loss into a unified picture of hydrodynamic escape is necessary. This modelling should help identify the regions of parameter space that each mass-loss mechanism dominates. Further, observations of atmospheric escape for the emerging class of very young planets that are the likely antecedents of mature sub-Neptune size planets \cite{david16, david19a, david19b, newton19, plavchan20, rizzuto20} offer the hope of distinguishing between the photoevaporative and core-powered atmospheric loss mechanisms. Ultimately, our quantitative insights into how these planets formed, such as the core-mass function and how much H/He these planets accreted, depend strongly on the assumed mass-loss model. 

The main uncertainty in our understanding of the formation of sub-Neptune systems is where large cores (planetary embryos) form. Do they originate past the snow line and undergo large-scale migration, or very close to their stars and only migrate to a limited extent? Future advances will likely be aided by a better understanding of the bulk compositions of close-in planets, in particular their volatile contents~\cite<e.g.>[]{Gupta2019,RogersOwen2020}. Improved observations and models of the structure and evolution of planet-forming disks will also play a role, as the disk determines how fast pebbles drift, where and when they accumulate to form planetesimals~\cite{drazkowska17}, and how fast and in what direction growing planets migrate~\cite{bitsch19}. 

Putting our Solar System -- and its lack of close-in super-Earths or sub-Neptunes -- in the context of extrasolar planets is a challenge~\cite<for a discussion, see>[]{raymond18}. Jupiter is the only Solar System planet that would be detectable if the Sun were observed with present-day technology. Understanding where our system fits within the bigger picture may thus rest on demographic studies that correlate the nature of inner and outer parts of planetary systems, including super-Earths and sub-Neptunes, Jupiter-like gas giants, ice giant analogs, and even debris disks~\cite{clanton16,suzuki16,barbato18,zhu18,bryan19,raymond11,moromartin15}. Fortunately, we are in a golden era of extrasolar planetary astronomy where the observational tools needed for these studies are rapidly advancing. The next 15 years are sure to bring dramatic surprises and insights to match those of the first 15 years of sub-Neptune planet discovery and characterization.

\acknowledgments
The data underling the previously published figures (Figures~\ref{fig:kepler}, \ref{fig:occurence}, \ref{fig:valley_comp}, \ref{fig:hot_Earths}, \& \ref{fig:k2-18}) are available in the corresponding publications. A comprehensive database of exoplanet parameters can be found at the NASA Exoplanet Archive (\url{https://exoplanetarchive.ipac.caltech.edu}). Data from this archive was used to create Figure~\ref{fig:arch}.\\

JLB acknowledges generous support over the years from NASA, the NSF, the David and Lucile Packard Foundation, the Heising-Simons Foundation, and the Sloan Foundation. SNR thanks the PNP program of the CNRS as well as the Agence Nationale pour la Recherche for funding of many of the ideas presented here (grant ANR-13-BS05-0003-002) and is grateful to all his colleagues involved in the {\em MOJO} project. JEO is supported by a Royal Society University Research Fellowship and a 2019 ERC starting grant (PEVAP). 


%
%

%
%
%
%
%

\end{document}